\documentclass{egpublarxiv}

\usepackage{egsr2019}
\usepackage[pdftex]{graphicx} 

\usepackage{soul}
\usepackage{amsmath}
\usepackage{amssymb}
\usepackage{microtype}
\usepackage{wrapfig}
\usepackage{xcolor}

\def\figurePath{./}
\def\myfigure#1#2{\begin{figure}[t]\centering\includegraphics*[width = \linewidth]{\figurePath#1}\vspace{-0.2cm}\caption{#2}\label{fig:#1}\end{figure}}

\def\mycfigure#1#2{\begin{figure*}[t]\centering\includegraphics*[clip, width = \linewidth]{\figurePath#1}\vspace{-0.2cm}\caption{#2}\label{fig:#1}\end{figure*}}

\def\mycfiguresize#1#2#3{\begin{figure*}[t]\centering\includegraphics*[clip, width = #3\linewidth]{\figurePath#1}\vspace{-0.2cm}\caption{#2}\label{fig:#1}\end{figure*}}

\newcommand{\eg}{e.\,g.,\ }
\newcommand{\ie}{i.\,e.,\ }
\newcommand{\etal}{et~al.\ }

\newcommand{\refSec}[1]{Sec.~\ref{sec:#1}}
\newcommand{\refFig}[1]{Fig.~\ref{fig:#1}}

\newcommand{\refTbl}[1]{Tbl.~\ref{tbl:#1}}

\definecolor{changecolor}{rgb}{.8,1,.7}

\newcommand{\change}[1]{#1}

\soulregister\ref7
\soulregister\cite7 
\soulregister\refFig7
\soulregister\cite7
\soulregister\ref7
\soulregister\pageref7
\soulregister\shortcite7
\soulregister\eg7
\soulregister\ie7
\soulregister\etal7

\DeclareGraphicsExtensions{.png,.jpg,.pdf,.ai,.psd}
\newcommand{\mysection}[2]{\section{#1}\label{sec:#2}}
\newcommand{\mysubsection}[2]{\subsection{#1}\label{sec:#2}}
\newcommand{\mysubsubsection}[2]{\subsubsection{#1}\label{sec:#2}}

\newcommand{\myparagraph}[1]{{\textbf{#1}}}

\SpecialIssuePaper
\PrintedOrElectronic

\usepackage{t1enc,dfadobe}

\usepackage{egweblnk}
\usepackage{cite}
\usepackage{booktabs}
\usepackage{microtype}
\usepackage{multirow}
\usepackage{flushend}
\usepackage{hyperref}

\title{Deep-learning the Latent Space of Light Transport}

\author[P. Hermosilla \& S. Maisch \& T. Ritschel \& T. Ropinski]
{
    \parbox{\textwidth}{
        \centering
        Pedro Hermosilla$^{1*}$ and
        Sebastian Maisch$^{1*}$ and
        Tobias Ritschel$^{2}$ and
        Timo Ropinski$^{1,3}$
    }
    \\
    \parbox{\textwidth}{
        \centering
        $^1$Ulm University, Germany \\
        $^2$University College London, United Kingdom \\
        $^3$Link{\"o}ping University, Sweden
    }
}

\begin{document}

\pagestyle{plain}

\teaser{
\includegraphics[width=.95\linewidth]{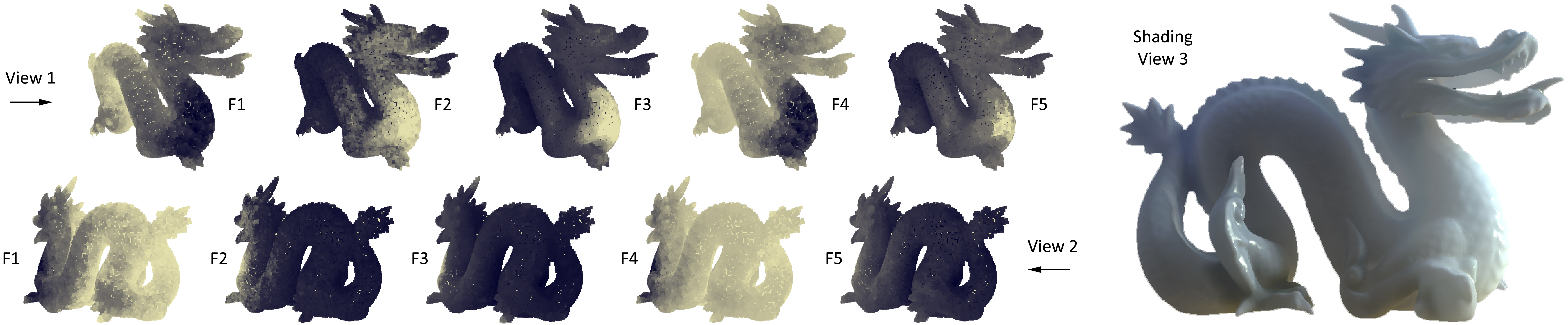}
\centering
\caption{We learn the latent space of light transport using CNNs operating on 3D point clouds. Here we show five resulting 3D feature activations  \emph{F1}-\emph{F5} for one geometry-light-material input \emph{(left columns)} from two views \emph{(left rows)} and the resulting shaded image \emph{(right)}.}
\label{fig:Teaser}
}

\maketitle

\begin{abstract}
We suggest a method to directly deep-learn light transport, \ie the mapping from a 3D geometry-illumination-material configuration to a shaded 2D image. While many previous learning methods have employed 2D convolutional neural networks applied to images, we show for the first time that light transport can be learned directly in 3D. The benefit of 3D over 2D is, that the former can also correctly capture illumination effects related to occluded and/or semi-transparent geometry. To learn 3D light transport, we represent the 3D scene as an unstructured 3D point cloud, which is later, during rendering, projected to the 2D output image. Thus, we suggest a two-stage operator comprising a 3D network that first transforms the point cloud into a latent representation, which is later on projected to the 2D output image using a dedicated 3D-2D network in a second step. We will show that our approach results in improved quality in terms of temporal coherence while retaining most of the computational efficiency of common 2D methods. As a consequence, the proposed two stage-operator serves as a valuable extension to modern deferred shading approaches.
\end{abstract}  

\footnotetext{* indicates equal contribution}
\footnotetext{This is the accepted version of the following article: \textit{Deep Learning the latent space of light transport}, which has been published in final form at http://onlinelibrary.wiley.com. This article may be used for non-commercial purposes in accordance with the Wiley Self-Archiving Policy http://olabout.wiley.com/WileyCDA/Section/id-820227.html.}

\mysection{Introduction}{Introduction}
The recent Artificial Intelligence (AI) break-through is also affecting image synthesis, with approaches that compute shading using networks~\cite{nalbach2017deep,kallweit2017deep}, for sampling~\cite{dahm2017learning,leimkuhler2018end,zheng2018learning,Mueller2018NIS,kuznetsov2018deep} and solutions that de-noise Monte-Carlo images~\cite{bako2017kernel,chaitanya2017interactive,vicini2018deep}. Realizing shading through a trained network enables it to become a building block in an AI ecosystem, \eg in deep inverse rendering~\cite{eslami2018neural,liu2018paparazzi,nguyen2018rendernet}, as every network always can trivially be back-propagated as part of another larger network. However, no method has yet been proposed to directly learn the mapping from a 3D scene description to a shaded image, which enables practical use in rendering.

Typically, convolutional neural networks (CNNs) operate on structured 2D images, and thus have found wide applications in all areas related to image processing. Regrettably, using 2D CNNs for shading, is inherently limited as information is incomplete and temporally unstable in 2D, as light transport lives in the space of mappings from a 3D representation to a 2D image. While CNN extension to regular 3D domains through voxel grids exist~\cite{wu2016learning}, they are too limited in resolution to resolve visual details, both for the input and output. For this reason, we suggest deep-learning the latent space of mappings from 3D point clouds to the resulting shading, which now becomes possible due to the recent progress in deep learning on unstructured point clouds~\cite{qi2017pointnet,hermosilla2018mccnn}.

To learn and use the latent space of light transport, the main challenge is to bridge the gap between the unstructured 3D scene description and the structured 2D output image. Therefore, we suggest a two-step operator, which allows for end-to-end learning from unstructured 3D to structured 2D images. The first step transforms the 3D input point cloud with per-point position, normals, material and illumination into a per-point latent feature vector, as it is visualized in~\refFig{Teaser}. This allows sharing information across space in multiple resolutions, a key feature of modern deep encoder-decoder learning architectures~\cite{ronneberger2015unet}. The obtained features are subsequently propagated onto all 2D pixels of the final output image in a second step, whereby we exploit an intermediate representation of these pixels as point cloud. Our evaluation finds improved quality and temporal stability when comparing the proposed approach with common 2D screen space solutions, a single-step variant or replacing each of the two steps by a non-deep encoding. We will demonstrate the quality of the proposed technique by training it to learn ambient occlusion (AO), global illumination (GI), and subsurface scattering (SSS) shading effects, which we evaluate quantitatively and qualitatively wrt.\ a path-traced reference.

\mysection{Related Work}{RelatedWork}
Our work addresses the problem of shading by combining principles from deep learning and point cloud geometry processing.

\paragraph*{Shading}
Computing an image from material, geometry and reflectance information is a core computer graphics problem with many proposed solutions.

Local illumination is fast but lacks support for complex shadows and lights or inter-reflections, but still dominates real-time and interactive applications such as computer games ~\cite{akenine2018realtime}. Physically-based ray-tracing~\cite{pharr2010physicaly} produces excellent  realistic results, but remains still too slow for dynamic scenes. It is typically used in combination with denoising-filters~\cite{bako2017kernel,chaitanya2017interactive,vicini2018deep} and temporal re-projection~\cite{nehab2007accelerating}, which each involve their own challenges. In contrast, we show how to obtain results that do not need either of the two.

Pre-computed radiance transport captures the light transport for a specific geometry and material such that illumination can be changed interactively~\cite{sloan2002precomputed}. This is feasible as light transport is linear in illumination but not in geometry~\cite{sloan2005local} or material~\cite{ben2008precomputed}. In our work, we do not use a neural network to replace a wavelet or Spherical Harmonics basis to code light transport for a single scene, which would be possible, given the ability to learn non-linear mappings, but methodologically rather under-ambitious. Instead, we learn the entire space of all light transports for all possible geometries, materials and illuminations.

In particular, our networks will learn a generalization of the direct-to-indirect transfer operator~\cite{hasan2006direct}. This operator maps the incoming direct light to the indirect light and previous work has compressed it or quickly evaluated it for a known scene (PRT: fixed geometry and material). Differently, we here learn a generalization that maps from the combination for dynamic direct light, dynamic geometry and reflectance to the indirect response.

Instant Radiosity~\cite{keller1997instant} comes probably closest to bridging offline and interactive rendering, but remains challenged by temporal coherence, singularities and resolving visibility between scene points and many virtual point lights (VPLs). Our approach uses point clouds that represent the first bounce, similar to VPLs, but we use a neural network to model their effect on the output image, which learns occlusion, multiple bounces and avoids the singularities resulting from an analytic derivation.

Our method can also be seen as a deep extension of point-based global illumination (PBGI)~\cite{christensen2010point,ritschel2009micro}. In those classic methods, last-bounce illumination (final gathering) is computed at 2D image positions, by gathering illumination from many 3D points.  This process can involve complex reflectance functions and most of all, visibility, which is found by splatting points into a small $z$-buffer. The relevant points are found in sub-linear time in PBGI. Our method has to learn those steps, including visibility. PBGI computes shading of $n$ pixels in respect to $m$ points in $\mathcal O(n\times\log(m))$ steps. We use an encoder-decoder \cite{ronneberger2015unet} that operates in $\mathcal O(n\times m)$, \ie time constant in the number of input points (like a MIP map it has a logarithmic number of levels, but exponentially fewer points at each level). Unfortunately, implementations of PBGI can be intricate making comparison difficult while ours is simpler to implement if a point-based learning framework is extended with our 3D-to-2D convolutions. Finally, PBGI cannot be back-propagated.

Screen space shading techniques exist for a multitude of effects, ranging from AO~\cite{mittring2007course}, over single-bounce diffuse GI~\cite{ritschel2009approximating} to SSS effects~\cite{jimenez2009screen}. These are fast approximations of the desired illumination effects, but unfortunately often lack accuracy or temporal stability. The reason for both is that information outside the image -- either due to being outside the viewport or occluded -- cannot be taken into account when computing the result. Mara~\etal\cite{mara2016deep} tackled these problems by using layered depth images (LDI)~\cite{shade1998layered}, which resolve occlusion but cannot represent geometry seen under oblique angles, outside the viewport and do not scale to scenes involving a high depth complexity.

\mycfigure{TrainingData}{
\protect\change{
Training data for different effects (AO, GI and SSS).
Three shapes are shown for each effect.
For GI and SSS, we also show the illumination used for each shape.
In each shape, we show the point cloud attributes that are input and output to the network.
We omit repeating the attributes of SSS that also go into GI for clarity.
Note that correlation in this figure is for illustration only; all data are randomized.
}
}

\paragraph*{(Deep) Learning}
Learning image synthesis is of interest to computer graphics and to vision, \ie inverse graphics~\cite{eslami2018neural}.

An early successful application of learning to a shading task is to regress radiance in a PRT setting \cite{ren2013radiance}. The idea is to compress the radiance response function using a fully-connected network. As all PRT work, this remains limited to compressing the light transport in one specific scene: instead of using SH or wavelets, the signal is compressed into a small per-vertex fully-connected network, that is efficient to evaluate for new view or illuminations. Follow-up work has applied the same idea to relighting of captured scenes~\cite{ren2015image}, where a small network encodes the response to illumination. To tackle learning the more challenging, high-dimensional space of light transport in all scenes, not just to compress low- dimensional response to light or view changes in a single specific one, requires more advanced DL concepts.

Nalbach~\etal\cite{nalbach2017deep} learn shading directly on a deferred framebuffer, as done in screen space shading. Their method is applicable to all deferred buffers, \ie it generalizes across scenes. However, it inherits the difficulties of screen space, and suffers from non-visible geometry, under-sampling and temporal instability. 

Kallweit~\etal\cite{kallweit2017deep} employ  deep learning for volumetric light transport. While they archive impressive results, their algorithm is not suitable for real-time applications and was only demonstrated to work on 3D volumes such as clouds. Their approach also operates in 3D where one network that probes the volume at multiple locations is used to regress shading in isolation for each location. We instead employ a single encoder-decoder architecture~\cite{ronneberger2015unet}, regressing shading at all pixels from a 3D point cloud as a whole. This approach increases efficiency and expressiveness using a latent space with spatially-shared internal features (\refFig{Teaser}).

Combining 3D with 2D information is also an important topic in computer vision. The work of Kar~\etal\cite{kar2018multiview} investigates transferring from images to a regular 3D voxel grid. Our work contributes a network layer that transfers from an irregular 3D  point cloud to the regular 2D image required for output instead.

\paragraph*{Point Clouds}
Point clouds are a minimal scene representation that do not contain any connectivity, but a raw sampling of the scene attributes in 3D~\cite{pfister2000surfels}. Originally, point clouds have been employed in geometry processing and have recently also become a subject of study for AI~\cite{qi2017pointnet,hermosilla2018mccnn}.

Several algorithms exist for computing global illumination effects in point clouds. Classic radiosity can be made meshless~\cite{lehtinen2008meshless}, operating on a point hierarchy. Bunnel describes a hierarchical technique for rendering AO and indirect illumination~\cite{bunnell2005dynamic}. More recently, point clouds have also been used in off-line rendering for movie production to approximate global illumination~\cite{christensen2010point}. While point clouds have many advantages as they are also the immediate output of acquisition devices, such as LIDAR scanners, direct, unstructured data like this can be hard to handle especially in real-time environments. For such cases, screen space algorithms have the advantage of a more spatially coherent memory structure.

In the rendering community, in particular for production rendering, point clouds with attributes are also known as Deep Framebuffers~\cite{Kainz2009EXR}. Deep Framebuffers have been proven useful also for shading~\cite{nalbach2014deep}. It is worth noting the difference of full deep framebuffers and Layered Depth Images (LDIs)~\cite{shade1998layered}. LDIs are regular and efficient, can resolve occlusion and transparency, but lack information outside the image and inherently cannot resolve geometry seen under oblique angles, which limits their use for shading~\cite{mara2016deep}. Therefore, our approach learns directly on a 3D point cloud instead of an LDI.

Most point-based methods need a way to propagate their 3D information to the final pixels. Splatting is popular~\cite{gautron2005radiance,scherzer2012preconvolved}, but can be difficult to adjust and is also computationally demanding. We learn this operation end-to-end with the 3D latent space.

\mycfigure{Overview}{
\protect
\change{
Overview of our approach.
The network receives as input two point clouds PC1 and PC2.
The first point, PC1, cloud is processed by our neural network to extract abstract features, which are used to compute AO/GI/SSS values for each point in PC2.
During training (red arrows, top branch), PC2 is another point cloud covering the surface of the object.
During testing (blue arrows, bottom branch), PC2 is the point cloud of 3D pixel coordinates.
Each box represents a Monte Carlo convolution in which the number of features (F) and radius (R) is indicated.
Blue boxes indicates a convolution on the same resolution of PC1 whilst green boxes indicates convolution between different resolutions. The orange box represents the Monte Carlo Convolution used to transfer features between PC1 and PC2.
}
}

\mysection{Learning Light Transport on 3D Point Clouds}{OurApproach}
The technique is composed of two parts: training data generation and a two-stage testing component (``Training'' and ``Testing'' in \refFig{Overview}).

\mysubsection{Training Data Generation}{TrainingData}

\paragraph*{Geometry Sampling}
To generate our training data, we use 200 random samples from SHREC~\cite{pickup2014shrec} and \change{800 from} Shapenet~\cite{chang2015shapenet}. To each of these meshes we assign a constant random material, before sampling them at \change{$n_\mathrm{3D}=20,000$} points using uniform random sampling. Note that this samples a 2D surface embedded in 3D, \ie it represents only the relevant part of the domain, while a 3D voxelization would require to represent the entire domain. Furthermore, all scene objects are re-scaled to have the same size to fit $[-1,1]^3$ and are placed on a ground plane.

\paragraph*{Shading Computation}
To compute the shading at each sample point as linear RGB radiance, we use a modified version of the Physically Based Rendering Toolkit (PBRT)~\cite{pharr2010physicaly}. During this computation, we send 128\,/\,4096\,/\,4096 rays when rendering AO\,/\,GI\,/\,SSS for each sample point such that the samples are reasonably noise-free. The shading is computed under illumination conditions, which are captured by 30 different environment maps \change{out of which a subset is shown in} \refFig{TrainingData}.

\paragraph*{Shading Effects}
To investigate a variety of shading effects, we have collected different variants of training data for the three desired shading effects: AO, GI and SSS.

For AO, positions and normals are stored as input, and we regress scalar gray AO as the shading variable. The ambient occlusion radius is chosen to be $.1$ of the scene radius.

For GI, we store position, normal, diffuse albedo, \change{(randomly uniform in $[0,1]$)} and direct illumination RGB irradiance as input, as well as the indirect RGB irradiance as output. This assumes both shadings to be direction-independent and leaves specular transport open for future work. We opt for irradiance as our output unit instead of radiant exitance, as it allows to include texture-modulated albedo details when converting from irradiance to pixel values, \ie display radiance. Here, we also include higher-order bounces, that are typically ignored in interactive global illumination.

Finally, SSS uses the same information as GI, just that the material information is extended by the reduced absorption coefficient that we choose as a random value from an exponential distribution with a mean at .1\,mm$^{-1}$ and the index of refraction chosen uniformly random between 1 and 1.5.

Please note, that all shadings are indirect illumination only, \ie our current use of the operator only includes the second and higher order bounces, but not the first bounce. We have chosen this proceeding, since very efficient specialized methods to compute this bounce exist, \eg (soft) shadow mapping. Future work could explore advanced direct shading effects such as complex arrangements of BRDFs, emitters and occluders.

\paragraph*{Split Protocol}
We split the generated data into a training data set of 20,000 point clouds (1,000 models $\times$ 20 environment maps), a validation data set with 1,000 point clouds (200 models $\times$ 5 environment maps) and a test data set with 2,500 point clouds (500 models $\times$ 5 environment maps).

Moreover, we also define an additional data set composed of animated 3D models of several animals. We use this data set to evaluate the ability of our learning algorithm to generalize, and to test the stability of the predicted values over an animation.

\mysubsection{Network Architecture}{Network}
As illustrated in \refFig{Overview} the proposed network architecture has two components. The first operates on the 3D point cloud itself (\refSec{3DStep}), the later propagates from the point cloud to the 2D image (\refSec{2DStep}). To bridge the gap between unstructured input data and a structured output, both stages are jointly trained end-to-end. As the structure of the network is the same for different shading effects, we describe the general structure here.

\mysubsubsection{Network Input}{Input}
\change{
The \emph{input 3D scene} (visualized in 2D for an elephant in \refFig{Overview}, left), is sampled to a n input \emph{3D point cloud} to cover the entire model's surface (\texttt{PC1} dots on the elephant).
Each 3D point, \ie sample, is labeled with the attributes required for the desired shading effect, \eg position, normal, materials.}

\change{
In addition to the point cloud, we require a second point cloud PC2 on which the shading is computed.
At training time (red arrows), \texttt{PC2} is just a different sampling of the surface.
At test time, \texttt{PC2} is formed by the 3D point cloud of all pixels visible in a 2D image.
}
All pixels in this image are labeled with the same kinds of attributes as the associated point cloud, \eg position, normal, materials. Please note, that the pixels, which live on a structured grid, usually do not form a subset of the unstructured 3D point cloud. While the 3D input encodes the scene and we construct deep features on it, only a small fraction of ``deep computation'' is actually done on the 2D image, which in the largest part is to define the desired output.

\mysubsubsection{3D Step}{3DStep}
To process the network's input, our approach exploits an unstructured deep network~\cite{qi2017pointnet} to map input attributes to deep per-point latent codes that can be used to shade a 2D image. In particular, we use an encoder-decoder architecture~\cite{ronneberger2015unet} with Monte Carlo (MC) convolutions~\cite{hermosilla2018mccnn}. As can be derived from the details in the Appendix~\refSec{MCConvolution}, this design is efficient to execute, and can deal with irregular sample distributions as required. 

In contrast to other unstructured learning approaches, the problem at hand requires us to bridge the gap between unstructured point cloud data and structured image data. 
This is achieved by making 3D convolutions efficient, carefully choosing the right 3D encoder-decoder, with the appropriate parameters and the insight that structured pixels in a 2D framebuffer are just a special case of an unstructured 3D point cloud. In the remaining paragraphs, we will elaborate on these design choices.

First, the input 3D points are resampled using a parallel Poisson disk sampling with a fixed radius on the $n_\mathrm{3D}$ points of the input of our data set, resulting in $n_\mathrm{3D}\prime$ points, which are used as the input of our network. Moreover, we now compute a point cloud hierarchy by repeatedly applying Poisson disk sampling with an increasing radius until we obtain a few points per model. The radii used to compute this hierarchy are $.01$, $.05$, $.15$, and $.5$. %Note, that they are defined relative to the size of the bounding box, and thus wrt. to the locality of the desired shading effects.

Naturally shading algorithms have to capture illumination effects at various different scales, reaching from local to global. To also capture this variety in our learning approach, we have chosen to follow the encoder-decoder architecture design~\cite{ronneberger2015unet}. The encoder processes each level of the point hierarchy by first applying a within-level convolution and then computing a convolution between the current level and the next one, which enables us to transfer the learned features to deeper levels. Before each convolution, we use a $1 \times 1$ convolution, \ie a  receptive field containing only the point itself, which is very fast to execute and allows to non-linearly adjust features as required. This procedure is executed for each level until we obtain a set of features for the last level of the point hierarchy. In each level the number of features is doubled, whereby we used in our particular implementation, 8 features in the first level which results in 64 features in the deepest level.

The decoder transfers the global features to shallow levels by applying convolution between different levels%. In the last level, we up-sample features from all the previous levels
, resulting in $n_\mathrm{3D}\prime \times n_\mathrm c$ channels in the last level (we use $n_\mathrm c$ equal to 8). We call this mapping $f_\mathrm{3D}$. Each 3D point $\mathbf x_i$ is now labeled with a latent encoding $\mathbf y_i\in\mathbb R^{n_\mathrm c}$ that describes its effect on shaded 2D pixels.

Our latent encoding could be seen as a deep generalization of VPLs~\cite{keller1997instant} or blockers and bouncers~\cite{sloan2007image} that encode what is relevant for other shading points. End-to-end training chooses this encoding optimally for the effect on the 2D image.

\mysubsubsection{2D Step}{2DStep} 
In this step, the latent 3D representation is propagated onto the 2D result image. Input are the  $n_\mathrm{3D}\prime$ points $\mathbf x_i$, with their $n_\mathrm c$ feature channels $\mathbf y_i$ as well as a large number of $n_\mathrm{2D}$ (\eg millions) of 2D image pixels with attributes $\mathbf z_i$. Output is a shaded image, \eg monochromatic for AO and RGB for other effects. Consistent with image-space CNNs, we use the term ``attribute" for given input information like position, normal, etc, while we refer to internal network activations, as ``features''. 

The propagation is performed using a single learned convolution that maps the $n_\mathrm c$ features as well as the per-pixel attributes of all points in a receptive field to a final RGB color, respectively a monochromatic gray value. We call this mapping, which is illustrated in \refFig{3DTo2D}, $f_\mathrm{2D}$.

\myfigure{3DTo2D}{Our key contribution is a learned convolution from sparse unstructured 3D features to dense 2D image pixels. A single \emph{(yellow point)}, collects information from 3D spatially nearby latent encoding of light transport \emph{(blue points)}. This is parallel to other pixels \emph{(gray points)} and independent of meshing \emph{(red lines)}.}

As this process is not creating any intermediate representations it is scalable. Note, that the number of 3D features is much smaller than the number of pixels. Our unstructured 3D-to-2D convolution matches those requirements. To determine all points $\mathbf x_i$ affecting a pixel $\mathbf z_i$, we look them up in a voxel hash map. Note that this map does not need to resolve fine spatial details but is just an optional acceleration data structure. In our implementation, we have chosen to use a $100^3$ grid, which resulted in adequate performance. A straightforward implementation of a single mapping from 3D to a 2D image in the spirit of MC convolutions~\cite{hermosilla2018mccnn} would require to build these structures for all pixels $\mathbf z_i$ in every frame, which would be prohibitive. Instead, our design allows hashing only the coarse point set $\mathbf x_i$. This strikes a balance between sharing information on a coarse scale and propagating this representation in a simple and scalable way to millions of pixels as required in practical computer graphics applications.

The ratio between the Poisson disk radius used to compute the $n_\mathrm{3D}\prime$ points and the radius of this last convolution determines the maximum number of points in the receptive field. Bounding the number of points used to compute the convolution allows us to guarantee a constant performance, since for each pixel $\mathbf z_i$ we will process a similar number of $\mathbf x_i$ points independently of the complexity of the scene. The trade-off between performance and quality of the effect can be controlled by increasing or reducing the Poisson disk radius. This will result in less or more points used to compute the convolution, obtaining a less or more accurate approximation of the integral. \refSec{Evaluation} presents a comparison of the results obtained for several scenes using different Poisson disk radii.

\mysubsection{Training Process}{Training}
During training, for each model we select the $n_\mathrm{3D}\prime$ points out of the initial $n_\mathrm{3D}$ points using Poisson disk sampling. The loss is $\mathcal L_2$ on their shading values. The remaining points are considered as the pixel points $\mathbf z_i$ for which we are approximating the shading effect. This is possible, as we interpret any pixel as a 3D point, entirely ignoring the image layout. Therefore, we are able to train our network end-to-end without generating several images from different points of view. Future work, could investigate the benefit of also using 2D images, \eg in an adversarial design, at the expense of having to render them.

The architecture is defined and trained using TensorFlow using the Adam optimizer at an initial learning rate of $.005$. We scaled the learning rate by $0.7$ every 10 epochs. The network is trained until convergence for 200 epochs using a batch size of 8 models.
\change{Test and train loss are similar, indicating no over-fitting is present.}
Our dataset and networks \change{are publicly available at }\url{https://github.com/viscom-ulm/GINN/}.

\mysubsection{Implementation}{Implementation}

\change{Our interactive OpenGL application proceeds as follows:}

\change{First, we compute a deferred shading buffer with position, normal, material and direct-light radiance, maybe with specular, using a vertex and fragment shader combination, all classic so far.}

\change{Next, we compute irradiance at every point of a point cloud version of this very scene, stored in a VBO, using compute shaders. This VBO, VBOs of position, normals and material information, are given to TensroFlow to compute the mapping $f_\mathrm{3D}$. The final 2D texture with radiant exitance is compute by a CUDA program which implements the mapping $f_\mathrm{2D}$ using as input the deferred position and normals of the pixels and the result of the mapping $f_\mathrm{3D}$.}

\change{Finally, this output composed of albedo, direct light, gamma applied, is tone-mapped and displayed.}

\change{Indeed, this requires having a point cloud version of the scene available.
In a pre-process, we sample the scene uniformly. However, under uniform polygonal tessellation, taking a random subset of the vertices should  be sufficient. }

\mysection{Evaluation}{Evaluation}
In this section we perform both a quantitative and time analysis (see \refSec{Quantitative}) and a qualitative evaluation (see \refSec{Qualitative}).

\mysubsection{Quantitative Evaluation}{Quantitative}
Here, we quantify different methods, including ours, variants of it, and other state-of-the-art methods in terms of several metrics on a test data set.

\mysubsubsection{Methods}{Methods}
For each of the three different shading effects considered (AO, GI, and SSS), we compare four different approaches to the reference: our full 3D-2D approach (Ours), conventional screen space techniques (SS), a 2D-only variant of our approach (Ours 2D only) and a 3D-only variant of our approach (Ours 3D only).

\paragraph*{Screen Space}
To obtain the screen space results, we use our own implementations of screen space shading, based on methods proposed for AO~\cite{mittring2007course}, GI~\cite{ritschel2009approximating} and SSS~\cite{jimenez2009screen}. For AO we sample 16 directions with 32 samples along each direction, which results in  512 samples per pixel. For GI and SSS we use a window of $54\times 54$, which results in $2,025$ samples per pixel.

\paragraph*{2D-only Variant}
For the 2D-only variant, we do not learn any 3D features per point, \ie we do not execute the encoder-decoder network. Instead, we learn only a single 3D-to-2D convolution which, based on the normal and other parameters of a sampling point, approximates the shading effects for each pixel. This can be understood as if we only execute the 2D part of our network, but with the same resources. Outperforming this method indicates, that sharing internal information in 3D is purposeful.

\paragraph*{3D-only Variant}
This ablation variant of our approach first computes the shading effects, \ie the RGB irradiance, at every input sample point. Recall, that the full approach does not do this, but creates a complex deep representation for every point instead. Then, we use a splatting technique to propagate the 3D irradiance onto the 2D image. Outperforming such a method would show that a 2D-3D design is advantageous over a pure 3D approach.

\mysubsubsection{Comparison Metrics}{Metrics}
To evaluate the shading methods, we compute measurements of all methods in comparison to a path-traced reference, whereby we employ three metrics. The first metric is computed in 3D, it is a direct view-independent $\mathcal L_2$ metric directly computed on the 3D point clouds. 
The second and third metrics are view-dependent as they are computed in 2D.
We use the mean square error of the resulting pixel values and the structural similarity (\change{DSSIM}) index computed on 2D images. For all these metrics, smaller values indicate better results.

\mysubsubsection{Additional 2D Test Data}{TestData}
The 3D metric can be evaluated directly on the 3D point clouds in the split set of our test data set. Recall, that the training operates purely on 3D point clouds, so we do not have a test image set, despite it is important to study the effect on the resulting image. To compensate for this, we rendered 5 additional reference 2D images with a resolution of $n_\mathrm{2D}=1024\times 1024$ for each shading effect and each method. All these images are linearly tone-mapped to preserves the .9 luminance percentile before applying a $1.6$ gamma curve. These images are shown in the supplementary materials.

\mysubsubsection{Timings}{Timings}
To gauge the performance of our approach, we also record the compute times \change{at a resolution of 1024$\times$1024} for each method on an Nvidia GTX 1080 / Intel i7-4790@3.60GHz system, and report it in \refTbl{Timing} together with the amount of memory used. Note that this operation is fully dynamic,
since the voxel grid is rebuild each frame in a few additional milliseconds. For static scene this could be pre-computed and stored in memory.
%with the only restriction that animations are isometric, \ie preserve distances. This is, as the voxel grid is build using geodesic distances, which do not change under deformations, allowing to share the grid across time. For scenes that drastically change shape, this voxel grid would need to be rebuild in a few additional milliseconds. 
We do not include the computation of direct illumination in time or memory consumption as this is completely independent of our approach or the methods we compare to. The 2D step of our network on average takes \emph{60\,\%} of the total compute time. However, depending on the scenes local-global characteristics, different weighting to 2D or 3D effort are possible.

Additionally, \refTbl{TimingPD} shows timings for different numbers of points, obtained by changing the Poisson disk radius. Larger radii result in larger receptive fields, \ie more points and more computational effort. We see that our method's compute time scales slightly sub-linear, almost linear in the number of points, indicating a controllable quality-performance trade-off.

\begin{table}[t]
\setlength{\tabcolsep}{1.5pt}
\caption{
Time and memory requirements for different methods (SS, 2D-only, 3D-only, Ours) \emph{(rows)} when realizing different shading effects (AO, GI, SSS) \emph{(columns)}.
}%
\label{tbl:Timing}%
\begin{tabular}{r rr r rr  r rr}
\toprule
&
\multicolumn{2}{c}{AO}&&
\multicolumn{2}{c}{GI}&&
\multicolumn{2}{c}{SSS}\\
\cmidrule{2-3}
\cmidrule{5-6}
\cmidrule{8-9}
&Time&Mem&&Time&Mem&&Time&Mem\\
\midrule
SS&
$1.8$\,ms&3.14\,MB&&$65.0$\,ms&9.34\,MB&&$31.0$\,ms&10.4\,MB\\
2D-only&
$26.5$\,ms&3.25\,MB&&$41.7$\,ms&9.61\,MB&&49.6\,ms&10.6\,MB\\
3D-only&
$71.8$\,ms&3.25\,MB&&$121.9$\,ms&9.61\,MB&&$104.0$\,ms&10.6\,MB\\
Ours&
$43.3$\,ms&3.25\,MB&&$107.6$\,ms&9.61\,MB&&105.6\,ms&10.6\,MB\\
\bottomrule
\end{tabular}
\end{table}

\begin{table}[t]
\setlength{\tabcolsep}{2.3pt}
\caption{
Number of points and time required to evaluate the network for different Poisson disk radii used to select the $n_\mathrm{3D}\prime$ points.
}%
\label{tbl:TimingPD}%
\begin{tabular}{rr r rr r rr  r rr}
\toprule
&&&\multicolumn{2}{c}{.01}&&\multicolumn{2}{c}{.015}&&\multicolumn{2}{c}{.02}\\
\cmidrule{3-5}
\cmidrule{7-8}
\cmidrule{10-11}
&&&\#Pts&Time&&\#Pts&Time&&\#Pts&Time\\
\midrule
\multirow{2}{*}{Elephant}&GI&&13.2\,k&280.1\,ms&&6.4\,k&122.9\,ms&&3.7\,k&81.2\,ms\\
&AO&&8.6\,k&203.6\,ms&&4.1\,k&66.1\,ms&&2.3\,k&34.6\,ms\\
\multirow{2}{*}{Horse}&GI&&9.0k\,&184.5\,ms&&4.4\,k&92.3\,ms&&2.5\,k&61.4\,ms\\
&AO&&3.5\,k&35.7\,ms&&1.7\,k&20.5\,ms&&0.9\,k&15.9\,ms\\
\bottomrule
\end{tabular}
\end{table}

\begin{comment}
\multirow{2}{*}{Eleph}&GI&&13,257&280.1\,ms&&6,411&122.9\,ms&&3,720&81.2\,ms\\
&AO&&8,604&203.6\,ms&&4,102&66.1\,ms&&2,385&34.6\,ms\\
\multirow{2}{*}{Horse}&GI&&9,095&184.5\,ms&&4,413&92.3\,ms&&2,569&61.4\,ms\\
&AO&&3,591&35.7\,ms&&1,713&20.5\,ms&&991&15.9\,ms\\

\end{comment}

\mysubsubsection{Discussion}{Discussion}
The results of our quantitative analysis shown in \refTbl{Results} demonstrate that our method (Ours) outperforms all other methods (SS, 2D-only, 3D-only) for all shading effects (AO, GI, SSS) according to all metrics (48 comparisons) with two exceptions. The first is computing GI in screen space, where SS, 3D-only and our full method perform similar ($.11$ vs.\ $.12$). The second is 3D-only, which is also our method, but an ablation. This indicates, that SSS, at least in our scenes, does not benefit from refining it from 3D to 2D, as this is the difference from 3D-only to our full method. We hypothesize, that this is due to the fact that SSS does not have high spatial frequencies which are worth to refine from 3D to 2D and the attempt to do so is counterproductive. Finally, while not outperforming state-of-the-art methods in all metrics for all effects, even at allegedly same performance, we find ours to have better temporal coherence, which is difficult to measure, but best seen in the accompanying video. Overall, we see that this increase in quality can require slightly higher compute time and only a very slight increase in memory by a few percent from \refTbl{Timing}.

In fact, we find that the ablations of our methods (2D-only and 3D-only) do not produce the same quality, given similar resources. We see this as an indicator, that our novel 3D-to-2D convolutional design, which bridges from unstructured to structured data is both, efficient and effective.

We further see, that the memory overhead is negligible as most processing happens on a light-weight point cloud of only ca.~10\,k points to capture global effects with a final 2D pass that needs an equal amount of memory as SS methods.

\mysubsection{Qualitative Evaluation}{Qualitative}

\refFig{AOResults} and \refFig{GIResults} show results for AO and GI for different methods (SS, Ours, Reference). We find that our method produces results that are more similar to the reference. The supplemental video further demonstrates our increased temporal coherence.

\change{We use point lights for testing GI, to highlight the effect of indirect lighting more. Note that this shows the networks ability to generalize to illumination not observed at training time. We additionally show results for environment maps in \refFig{EnvMaps}.
}

\refFig{sss} provides some examples of the subsurface scattering results obtained with our network. Note how our network is able to simulate back scattering effects.

\refFig{Ablation} shows a visual comparison of the obtained AO results when using the different methods: SS, 2D-only, 3D-only, Ours, Reference (from left to right). It can be seen, that SS (first inset column) mostly resolves local features. Our 2D-only variant (second inset column) has a similar quality, indicating that the 2D operation can be learned. The 3D-only variant of our approach (third inset column) in contrast resolves more global features, but lacks detail. When instead using our full method (fourth inset column), both local and global features are resolved, which makes it look most similar to the reference (fifth inset column).

In \refFig{PDRadiusComparison} we study the visual effect of different Poisson disk radii. We see, that with smaller radii, more 3D points map to every 2D pixel. Consequently, the shading appears more smooth, while still communicating details correctly.

\mycfiguresize{AOResults}{Results of different AO  methods \emph{(rows)} applied to different scenes \emph{(columns)}. The rendered meshes are not part of the training data set. \refTbl{Timing} and \refTbl{Results} provide a quantitative comparison.}{.95}

\mycfiguresize{GIResults}{Results of different GI  methods \emph{(rows)} applied to different scenes \emph{(columns)}. The rendered mesh-material combinations are not part of the training data set. \refTbl{Timing} and \refTbl{Results} provide a quantitative comparison.}{.95}

\mycfigure{Ablation}{Comparison of our method \emph{(left, full image)} with different methods: SS, 2D-only, 3D-only, Ours, Reference \emph{(insets left to right)}. As can be seen, SS and 2D-only capture local effects but lack global transport, while 3D-only captures global effects but lacks locality. Ours jointly learns both end-to-end, and allows us to obtain results similar to the reference. The shown mesh is not part of the training data set. Please refer to \refTbl{Timing} and \refTbl{Results} for quantification.}

\mycfiguresize{EnvMaps}{Global illumination (\textbf{middle}) results of our network for different environment maps. We also present the direct illumination (\textbf{left}) and the indirect illumination (\textbf{right}) separately for illustrative purposes.}{.95}

\mycfigure{sss}{
Subsurface scattering results of our neural network for different materials using environment maps (\textbf{left}), and point lights (\textbf{right}).}

\mycfiguresize{PDRadiusComparison}{
Obtained results for GI (top) and AO (bottom) for different Poisson disk radii used to select the representative points during testing. From left to right, .01, .015, and .02. Smaller circles result in more points contributing, which produces smoother effects as seen in the insets. In all results throughout this paper, we use $r=.15$ unless stated otherwise.}{.95}

\begin{table}[t]
\setlength{\tabcolsep}{1.7pt}
\caption{
Visual fidelity metrics computed wrt.\ to the reference in 2D and 3D (MSE, \protect\change{DSSIM}) for different methods (SS, 2D-only, 3D-only, Ours) \emph{(rows)} computing different effects (AO, GI, SSS) \emph{(columns)}.
Entries are shown to fall below two rows to highlight they are identical by construction: 3D is computed in 3D per-point and consequently not affected by our 3D-to-2D refinement.
}%
\label{tbl:Results}%
\begin{tabular}{r r rrr r rrr r rrr}
\toprule
&&
\multicolumn{3}{c}{AO}&&
\multicolumn{3}{c}{GI}&&
\multicolumn{3}{c}{SSS}\\
\cmidrule{3-5}
\cmidrule{7-9}
\cmidrule{11-13}
&&
\multicolumn{1}{c}{3D}&\multicolumn{2}{c}{2D}&\phantom{A}&
\multicolumn{1}{c}{3D}&\multicolumn{2}{c}{2D}&\phantom{A}&
\multicolumn{1}{c}{3D}&\multicolumn{2}{c}{2D}
\\
&&{\tiny MSE}&{\tiny MSE}&{\tiny \change{DSSIM}}&&{\tiny MSE}&{\tiny MSE}&{\tiny \change{DSSIM}}&&{\tiny MSE}&{\tiny MSE}&{\tiny \change{DSSIM}}\\
\midrule
SS&&
\multicolumn{1}{c}{--}&.24&.013&&
\multicolumn{1}{c}{--}&\textbf{.11}&\textbf{.041}&&
\multicolumn{1}{c}{--}&.49&.034\\
2D-only&&
.086&.25&.015&&
.062&.15&.043&&
.0215&.43&.018\\
3D-only&&
\multirow{2}{*}{\textbf{.073}}&.21&.014&&
\multirow{2}{*}{\textbf{.047}}&.12&.043&&
\multirow{2}{*}{\textbf{.0164}}&.29&\textbf{.013}\\
Ours&&
&\textbf{.16}&\textbf{.012}&&
&.12&.042&&
&\textbf{.19}&.017\\
\bottomrule
\end{tabular}
\end{table}

\mysection{Discussion}{Discussion2}
\change{
Our method is demonstrated in a setting were several assumptions are made: a single object of two diffuse and homogeneous materials,
For deployment to real rendering applications like computer games, several limitations would need to be overcome. 
}

\myparagraph{Diffuse materials}
\change{
It currently only works with uniformly distributing reflections (\ie diffuse materials). The largest difficulty to overcome for specular is the need to store directional illumination information at each point. A simple solution to start with would be using Spherical Harmonics, but these remain limited to low frequencies. Light transport is linear in light, so one could learn a isolated direction-dependent family of transports, but this is suboptimal from a deep learning perspective, as it would not allow for sharing internal features across the directional domain. 
}

\myparagraph{Homogeneous materials}
\change{
While our results are not shown textured, including reflectance variation would require mapping texels, including proper minification, to the coarse point's albedo feature which we did not implement. The 3D-to-2D step includes the reflectance of every pixel, that might or might not come form a texture. This is possible by learning irradiance to be multiplied with albedo, instead of radiant exitance.
}

\myparagraph{Single objects}
\change{
We only test and train on single object with one material placed on a ground plane of a different material. This is a constrained subset of what actual geometry would look like in many interactive applications.
Going from objects to scenes will face similar challenges that were encountered when using PRT.
}

\myparagraph{Computational efficiency}
\change{
Our approach is competitive in visual performance, but not yet able to outperform well-developed screen space interactive GI methods. We particularly note, how AO works better than SSS, which again works better than GI, probably as the these effects are increasingly demanding in reproducing high frequencies. Most time is spend by propagating from 3D points to 2D pixels, which every direct-to-indirect method needs to do.
}

\myparagraph{Scalability}
\change{
Our method scales linearly with the number of pixels: At each pixel, the 3D-to-2D convolution is run in isolation once. Future work can aim to reduce constants by using less, but better-trained filters. As an encoder-decoder, the 3D-convolution part scales linearly in the number of 3D points.
}

\mysection{Conclusions}{Conclusions}
In this work, we have proposed a deep learning approach to compute shading in its natural context: the full 3D geometry-material-illumination configuration. We could show how this can be achieved by extending modern scalable convolutional architectures, that directly work on the unstructured 3D scene sampling data. To our knowledge this is the first approach applying learning for rendering directly in 3D space. Our results show that we can outperform state-of-the-art deferred shading methods, as we consider parts of the geometry invisible to these. Thus, we believe that the presented approach is a valuable extension for these commonly used approaches.

Besides adding more effects, such as specular transport and testing the design on volumes, future work could also extend the approach to 4D, using temporal features, maybe including recursion to further increase efficiency. Inverting the pipeline and regressing the 3D information from the observed shading -- 3D intrinsic images -- is another avenue enabled by our approach.

\noindent
\textbf{Acknowledgements}
This work was partially funded by the Deutsche Forschungsgemeinschaft (DFG) under grant RO 3408/2-1 (ProLint), and the Federal Minister for Economic Affairs and Energy (BMWi) under grant ZF4483101ED7 (VRReconstruct). We would like to acknowledge the NVIDIA Corporation for donating a Quadro P6000 for our training cluster, and Gloria Fackelmann for providing the voice over the supplementary video.

\bibliographystyle{eg-alpha}

\bibliography{paper}

\appendix

\mysection{Monte Carlo Convolution}{MCConvolution}
\small
Monte Carlo convolutions (MCCs) \cite{hermosilla2018mccnn} are a deep neural network layer that efficiently convolves unstructured samplings of a signal $f$ with a learnable kernel $g$:
\begin{equation}
(f\ast g)
(\mathbf x)
\approx
\frac{1}{|\mathcal N(\mathbf x)|}
\sum_{j\in\mathcal N(\mathbf x)}
\frac{
f(\mathbf y_j)g\left(\frac{\mathbf x - \mathbf y_j}{r}\right)
}{
p(\mathbf y_j|\mathbf x)
},
\label{eq:MCConvPDF}
\end{equation}
where $\mathcal N(\mathbf x)$ is the set of all samples in the neighborhood of spatial coordinate $\mathbf x$, $p(\mathbf y|\mathbf x)$ is the density around sample $\mathbf y$ in respect to point $\mathbf x$ and $r$ a scalar defining the radius of the receptive field.
The kernel $g$ takes as arguments the 3D offsets $\mathbf x-\mathbf y_i$, and maps them to weights, same a discrete filter masks do in an image filter.
Learning the filter amounts to learning the weights of the MLP defining it.
To be applicable to high dimensions, MCCs model the filter kernel itself as a Multi-layer Perceptron (MLPs), a network which map 3D offsets to scalar weights.
As the definition allows completely decoupling the input and output sampling, MCCs are well-suited to down and up-sampling, as well as to the change of dimension from 3D virtual worlds to 2D image pixels required here.

\end{document}